\documentclass{article}
\usepackage[utf8]{inputenc}
\usepackage[english]{babel} % Включаем пакет для поддержки русского языка.
\usepackage[dvips,pdftex]{graphicx}
\usepackage{float}
\usepackage{perpage}
\usepackage{amssymb}
\usepackage[section] {placeins}
\usepackage{authblk}
\usepackage[round,authoryear]{natbib}
\MakeSorted{figure}
\MakeSorted{table}
\title{The K-corrections to radial velocity curves of optical components
in X-Ray Binaries. HMXB with weak X-ray heating.}
\date{26.12.2012}
\author{V. S. Petrov \footnote{e-mail: vpetrov@sai.msu.ru}, E. A. Antokhina, and  A. M. Cherepashchuk}
\affil{\textit{Sternberg Astronomical Institute, M.V. Lomonosov Moscow State University}}
\begin{document}
\maketitle
\begin{abstract}
The radial velocity curves of optical components in X-ray binary systems can differ from the radial velocity curves of their barycenters due to tidal distortion, gravitational darkening, X-ray heating, etc. This motivated us to investigate how the semiamplitudes of the radial velocity curves of these optical components can depend on the binary-system parameters in a Roche model. The K-correction is taken to be the ratio of the radial velocity semiamplitude for a star in the Roche model to the corresponding value
for the stellar barycenter. K-corrections are tabulated for the optical stars in the massive X-ray binaries Cen X-3, LMC X-4, SMC X-1, Vela X-1, and 4U 1538-52.

\end{abstract}
\section{INTRODUCTION}
Accurate determination of the masses of the compact objects in X-ray binaries remains a topical problem in modern astrophysics. Although current astronomical facilities can be used to obtain high-accuracy spectra and radial velocity curves of binary stars, fitting
these data correctly remains challenging. Observed stellar radial velocity curves can display systematic distortions due to tidal interaction between the binary components, X-ray heating effect, gravitational darkening, etc. However, observed stellar radial velocity curves are often analyzed in models with two point masses, when the shape of the curve does not depend on the nearness of the components. The K-correction to the semiamplitude of the radial velocity curve is introduced to partially take into account
effects that arise when the components are close to one another.

The effect of the asphericity of a star in a binary
system on its radial velocity curve was first considered 
in 1941 by \citet{ss1}, who analyzed the tidal
deformations of stars in spectroscopic binaries as a source of false eccentricities. Stern derived an analytical expression for a correction to the radial velocity of a deformed star, taken to be an ellipsoid of rotation with gravitational darkening. Later, analytical formulas for corrections to observed radial velocities were
derived by \citet{ss2,ss3,ss4}. 

 \citet{ss7} was the first to consider the effect of tidal deformations of stars when calculating line profiles and radial velocity curves using synthesis methods. A spherical star and a star in a Roche model were considered. The local profiles of area
elements were taken to be either theoretical profiles with simple (Gaussian) forms or observed profiles from a standard-profile library without considering the stellar rotation. This method was applied to several massive X-ray binaries and enabled estimation of their parameters.

 \citet{ss5} proposed direct calculation of radial velocity curves using a synthesis method (analogous to the method of \citealt{ss6}). The star was considered using a Roche model. The mean effective velocity of the visible stellar
disk relative to the stellar barycenter was calculated
using the formula
\begin{eqnarray}
\Delta V &=& \frac{\int vF dS}{\int F dS}
\end{eqnarray}
where $F$  is the flux from an area element in the direction toward the observer, $v$ is the radial velocity of the element relative to the stellar barycenter, and $dS$ the area of the surface element.  Appreciable deviations from a point mass model were noted when the star maximally filled its Roche lobe and the component mass ratio $q$ in the binary system was very small (the star is appreciably more massive than the compact object). Both features are characteristic of massive X-ray binaries with neutron stars.

 \citet{ss16} proposed an algorithm for calculating the line profiles and radial velocity curves of tidally deformed stars in close binary
systems using a synthesis method. The stars were treated using a Roche model, with their orbits being either circular or elliptical.Theoretical hydrogen line profiles \citet{ss22}  for various effective temperatures and surface gravities were adopted as the local line profiles of area elements. \citet{ss13} used this algorithm to model line profiles in close binaries, and concluded that the component mass ratios $q$ and the orbital inclinations $i$ could be independently determined using the variations of the stellar line profiles during the orbital period. A similar method for determining $q$ and $i$ was independently proposed by  \citet{ss14}. \citet{ss15} estimated the orbital inclination of the X-ray binary Cyg X-1 by applying the calculation algorithm of  \citet{ss16} to a high accuracy observed radial velocity curve, without using any light-curve data.

Later,  \citet{ss9} improved this algorithm
for the synthesis of line profiles and radial velocity curves of stars in X-ray close binaries. The main difference from previous versions of the algorithm \citet{ss16} concerns the calculation of the line profiles of the local areas and the treatment of X-ray
heating. The optical star is treated using a Roche model. The heating of the stellar surface by incident X-ray radiation from the relativistic object is included,
and the emergent radiation calculated by solving the radiative transfer equation at a given point in the stellar atmosphere. The new algorithm \citet{ss9} makes
it possible to take into account more correctly X-ray heating of the star by its companion, and to explain spectroscopic observations of close binaries.

 \citet{ss25,ss26,ss27,ss28,ss29}  used the algorithms
of  \citet{ss16,ss9} to analyse the radial velocity curves of some massive and low-mass X-ray binaries, refining the masses of these stars and compact objects.

The methods listed above enable the direct calculation of the line profiles and radial velocity curves of close binaries using sophisticated models. However, radial velocity curves are often fitted using models with two point masses.  \citet{ss18} introduced the idea of a “K-correction” related to the
difference between the radial velocities of a stellar barycenter and the “effective center” of a region where the spectral lines are formed. This makes it possible
to consider the effects of the nearness of components in a first approximation, without direct calculation of the radial velocity curves in complex models
(Roche models, models with rapidly rotating stars, etc.). Tabulated values of calculated K-corrections can be used to correct the semiamplitudes of radial velocity
curves in point-mass models. 

K-corrections are often used to correct the semiamplitudes
of radial velocity curves of optical stars in low-mass X-ray binaries. In these binaries, the lines of the optical components are very weak, since the contribution of the accretion disks dominate the total luminosities of the systems, and the regions of line formation on the optical stars can be shifted. For example, \citet{ss17}  found a systematic discrepancy between the radial velocity semiamplitudes of a late-type optical star in outburst and the quiescent state. This was explained as a result of heating of the stellar surface by radiation from the accretion disk, leading to a shift of the region where the absorption lines are formed.
 \citet{ss19} calculated K-corrections for the normal
star in the binary system IX Vel for two different accretion disk opening angles and several orbital plane inclinations in a model taking into account the geometry and kinematics of the accretion disk. 

The most complete study of radial velocity K-corrections
for stars in low-mass X-ray binaries to date was carried out by \citet{ss20}. They distinguished the Bowen emission lines \citet{ss21}, from
the spectrum of the system X1822-371 (V691 CrA), which they used to construct the radial velocity curve for the optical star. Tables of K-corrections for various
component-mass ratios and accretion-disk opening angles are provided in \citet{ss20}.
 
\section{FORMULATION OF THE PROBLEM}

The goal of our study is tabulating K-corrections based on direct calculations of theoretical stellar radial velocity curves. We have calculated radial velocity
curves for a range of parameters characteristic of massive X-ray binaries. We plan to carry out calculations for low-mass X-ray binaries in a future paper.

We have considered a massive X-ray binary system in a Roche model, varying the component mass ratio in the binary $q=M_x/M_{v}$, the effective temperature of the optical component $T_{eff}$, the Roche lobe filling factor of the optical star $\mu$ and the orbital inclination $i$. The remaining parameters of the binary were fixed. Here, $M_x$ is the mass of the X-ray component and $M_v$ the mass of the optical star. The model
radial velocity curves were calculated using the $H_{\gamma}$ line. The semiamplitudes and corresponding K-corrections were determined for the model radial velocity
curves, calculated as the ratio 
\begin{eqnarray}\label{eq:Kcorr}
K_{corr} &=& \frac{K^{Roche}_{v}}{K^{c}_{v}}
\end{eqnarray}
where $K^{Roche}_{v}$ is the maximum semiamplitude of the
stellar radial velocity curve in the Roche model and  $K^{c}_{v}$ the semiamplitude of the radial velocity curve of the stellar barycenter. Note that $K^{Roche}_{v}$ does not necessarily correspond to one of the quadratures (orbital
phases $\phi_{orb}=0.25$ and $0.75$).

X-ray heating was assumed to be either small ($ k_{x} = L_{x}/L_{opt} \leq 2$) or absent. Here, $k_{x}$ is the ratio of the X-ray luminosity of the relativistic component to
the bolometric luminosity of the optical star. If the X-ray heating is weak, it can be taken into accounted in a model in which the bolometric fluxes of the incident
X-ray radiation and the radiation of the optical star are simply added, without considering the transfer of the incident radiation in the atmosphere. 

We calculated the values of K-corrections for the five X-ray binaries with OB supergiants Cen X-3, LMC X-4, SMC X-1, Vela X-1, and 4U 1538-52, given in summary tables in the Appendix. These tables may be helpful when determining the masses
of X-ray pulsars (for example, using Monte Carlo simulations). 

This technique (e.g. \citealt{ss10}) usually assumes that the observed
radial velocity semiamplitude of the optical star, $K_{v}$ is given, so that it can be fixed. However, if the masses of the optical star and the relativistic object
were determined using Monte Carlo simulations in a point mass model, the observed value of $K_{v}$ must be corrected for the ellipticity and reflection effects. When varying $q$, $\mu$ and $i$, each set of these parameters is associated with its own correction factor. Therefore, we must use a specific value of the “observed” radial velocity semiamplitude for the optical star at each iteration step in the Monte Carlo method:
\begin{eqnarray}\label{eq:Kv}
K^{c}_{v}(q,\mu,i) &=& \frac{K_{v}}{K_{corr}(q,\mu,i)}
\end{eqnarray}
Here, we used the observed radial velocity semiamplitude,$K_{v}$ , instead of the quantity $K^{Roche}_{v}$ from (\ref{eq:Kcorr}). The corrected, observed stellar radial velocity is
referenced to its barycenter, which corresponds to the point-mass model. The actual observed value of $K_{v}$  depends on the nearness of the binary components. The fact that we used the refined value $K^{c}_{v}(q,\mu,i)$ instead of  $K_{v}$ ensures the correctness of the point-mass model applied to the X-ray binary being analysed.

On the other hand, the observed value of $K_{v}$ can be
compared with the theoretical radial velocity semiamplitude, $K^{Roche}_{v}$, calculated in the Roche model. We present tabulated K-corrections $K_{corr}(q,\mu,i)$ for
each of the systems analyzed in \citet{ss10}. With these, the theoretical radial velocity semiamplitudes $K_{point}$ calculated in the point-mass model
can be corrected for the ellipticity and reflection effects in each iteration step in the Monte Carlo method:
\begin{eqnarray}\label{eq:Kth}
K^{Roche}_{v}(q,\mu,i) &=& K_{point}(q,\mu,i)\cdot K_{corr}(q,\mu,i)
\end{eqnarray} 
$K^{Roche}_{v}(q,\mu,i)$ is then compared with the observed semiamplitude $K_{v}$. Thus, the corrected value $K_{point}(q,\mu,i)$ is used instead of the observed value $K_{v}$, which is distorted due to effects arising due to the nearness of the components

Note that only the semiamplitude of the observed radial velocity curve is considered in the Monte Carlo method; information about the shape of this curve is lost. A more correct analysis would consider the shape of the entire radial velocity curve, leading to
more reliable mass estimates for X-ray pulsars in binaries with OB supergiants (see, e.g., \citet{ss15})

Therefore, our tabulated K-corrections can be helpful in studies aiming to realize a more correct application of the Monte Carlo method to determine the masses of close binary components.
\section{THE BINARY MODEL}
We calculated theoretical profiles of the absorption lines and radial velocity curves of optical stars in X-ray binary systems using two algorithms, described in detail by  \citet{ss16, ss9, ss11}. We will refer to these as Algorithm I (\citealt{ss16, ss11}) and Algorithm II (\citealt{ss9}). The only difference between them is the means of
calculating the local line profiles for area elements on the stellar surface, which are then used to calculate the integrated line profile of the whole star.

Algorithm I uses profiles of the Balmer absorption
lines for various effective temperatures and surface gravitates calculated and tabulated by \citet{ss8}.In Algorithm II, the emergent radiation fluxes of the area elements and the line profiles are calculated via atmospheric modelling, with and without an external X-ray flux. We provide a brief description of the model
and calculation technique below.

A binary system model, consist of an optical star and a point-like relativistic object moving in circular or elliptical orbits around the system barycenter. The orbital plane is inclined to the plane of sky by an angle $i$. The component mass ratio is $q=M_x/M_{v}$, where $M_x$ is the mass of the compact object and $M_{v}$ the mass of the optical star. The optical star has a tidally deformed shape and inhomogeneous temperature
distribution over its surface due to gravitational darkening and X-ray heating by the radiation from the companion. The shape of the star coincides with the equipotential surface in the Roche model \citet{ss3, ss12}. The geometric size of the star is determined by the critical Roche lobe filling factor $\mu=R/R^*$, i.e., the ratio of the polar radii for a star that can partially or fully fill its critical Roche lobe at the orbital periastron \citet{ss11}. The star rotates asynchronously with the orbital revolution. The degree of asynchrony of the rotation is defined by the parameter $F=\omega_{rot}/\omega_K$, where $\omega_{rot}$ is the angular velocity of the star and, $\omega_K$ the mean Keplerian orbital angular velocity ($\omega_K=2\pi/P$,  where $P$ is the binary period).

The tidally deformed stellar surface was divided into area elements, for each of which the intensity of the emergent radiation was calculated. The calculated fluxes take into account gravitational darkening, heating of the stellar surface by incident radiation from the companion (the reflection effect), and limb darkening. In Algorithm I, the absorption profile and its equivalent width were calculated for
each visible area element, with the temperature $T_{loc}$ and local surface gravity $g_{loc}$ interpolating the tables of Kurucz for Balmer lines \citet{ss3}. The heating of the stellar surface by the X-ray radiation from the companion was included by adding the incident and emergent radiation, without considering radiation transfer in the stellar atmosphere. This approach is not fully correct; in particular, it does not take into account line emission that can arise when the incident radiation strongly heats the atmosphere. This simplified model of the reflection effect can be used only for X-ray
binaries with weak X-ray heating, $ k_{x} = L_{x}/L_{v} \lesssim 2$, where $L_x$ and $L_v$ are the bolometric luminosities of the X-ray source and the optical star, respectively.

In Algorithm II (\citealt{ss9}), an atmosphere model was
constructed at specified points of the stellar surface, to calculate the local line profile for each area element.The spectrum of the external radiation from the compact
source was specified based on X-ray observations or using a model function.
Apart from the local temperature $T_{loc}$ and the local surface gravity $g_{loc}$, the
parameter $k_x^{loc}$, equal to the ratio of the incident X-ray flux and the flux of the emergent radiation, was calculated for each area element, without considering the
external irradiation of the atmosphere. Using these parameter values at a specified point of the surface, the model atmosphere was calculated by solving the equations of radiative transfer in a line, including the effect of the incident external X-ray flux \citet{ss9}. Further, the emergent radiation intensities in the line and continuum were calculated using the adopted model atmosphere for each area element.

The local line profiles calculated using Algorithm I or II were added over the visible stellar surface, with allowance for the Doppler effect, after first normalizing
to the continuum for each area element. In this way, the integrated line profile from the star for a given orbital phase was calculated. The calculated integrated line profiles were used to determine the stellar radial velocities (for more detail, see \citealt{ss11, ss16}). The advantages of Algorithm I are its simplicity and computational speed; however, it can only be used when the X-ray heating is not strong. Algorithm II
is considerably more expensive in terms of computing time, but can be used to calculate the spectral line profiles of tidally deformed stars in X-ray binaries fairly correctly. Earlier modelling based on Algorithm II \citet{ss9} has shown that, in the presence of appreciable X-ray heating, absorption line profiles can be considerably distorted due to an emission component whose intensity varies with the orbital phase. This emission component can appreciably affect radial velocities derived from the line profiles of the optical star, which must be taken into account when analysing the
radial velocity curves. 

In this study, we restricted our consideration to X-ray close binaries with weak X-ray heating. A forthcoming paper will be devoted to calculations of radial velocities and K-corrections for X-ray close binaries with appreciable X-ray heating. 

\section{MODEL CALCULATIONS}

In the first stage, we carried out the above procedures and studied the K-corrections as functions of the binary parameters given in Table \ref{tabular:model}. We examined
the effects of several parameters: the component mass ratio $q=M_x/M_v$, the orbital inclination $i$, the Roche lobe filling factor $\mu$, and the gravitational darkening coefficient $\beta$. When varying one of the listed values, the other model parameters were held fixed. 

The K-correction was calculated using the formula $K_{corr} = K^{Roche}_{v}/K^{c}_{v}$, where $K^{Roche}_{v}$ is the maximum semiamplitude of the stellar radial velocity curve
in the Roche model and $K^{c}_{v}$ is the radial velocity semiamplitude of the stellar barycenter. Note that the maximum semiamplitude of the stellar radial velocity
curve in the Roche model $K^{Roche}_{v}$ does not necessarily correspond to one of the quadratures. 

We first modeled the K-corrections as functions of $q$ and $\beta$. Varying the gravitational darkening coefficient $\beta$ makes it possible to study the effect of the
brightness distribution over the visible stellar disk on the radial velocities. The coefficient $\beta$ determines the temperature of a stellar surface element: $T = T_{eff}(g/\bar{g})^{\beta}$, where $g$, $\bar{g}$ are the local and mean surface gravitaties, respectively, and $T_{eff}$ is the mean effective temperature of the star. According to Von Zeipel’s theorem, $\beta = 0.25$ for stars with radiative energy transfer \citet{ss30}. Empirical values of $\beta$ for stars in radiative equilibrium are provided in various other studies. For example, \citet{ss23} obtained $\beta = 1.36 \pm 0.04$, although this result was not confirmed in \citet{ss24}.   In our modeling, $\beta$ was varied between 0 and 0.5. The case $\beta =0$ corresponds to a homogeneous temperature distribution over the stellar surface, while the temperature of the area elements
can strongly vary depending on their localizations when $\beta=0.5$. If the star nearly fills its Roche lobe and there is no X-ray heating, the coolest part of the star is near the inner Lagrangian point, where the temperature can differ significantly from the polar
temperature. For example, for $q = 1$, the temperatures vary from 24 000 K to 12 000 K between the pole of the star and the region close to the inner Lagrangian point. The mean effective temperature is $T_{eff}=22000$~K.

\begin{figure}[h]
\center{\includegraphics[width=1\linewidth]{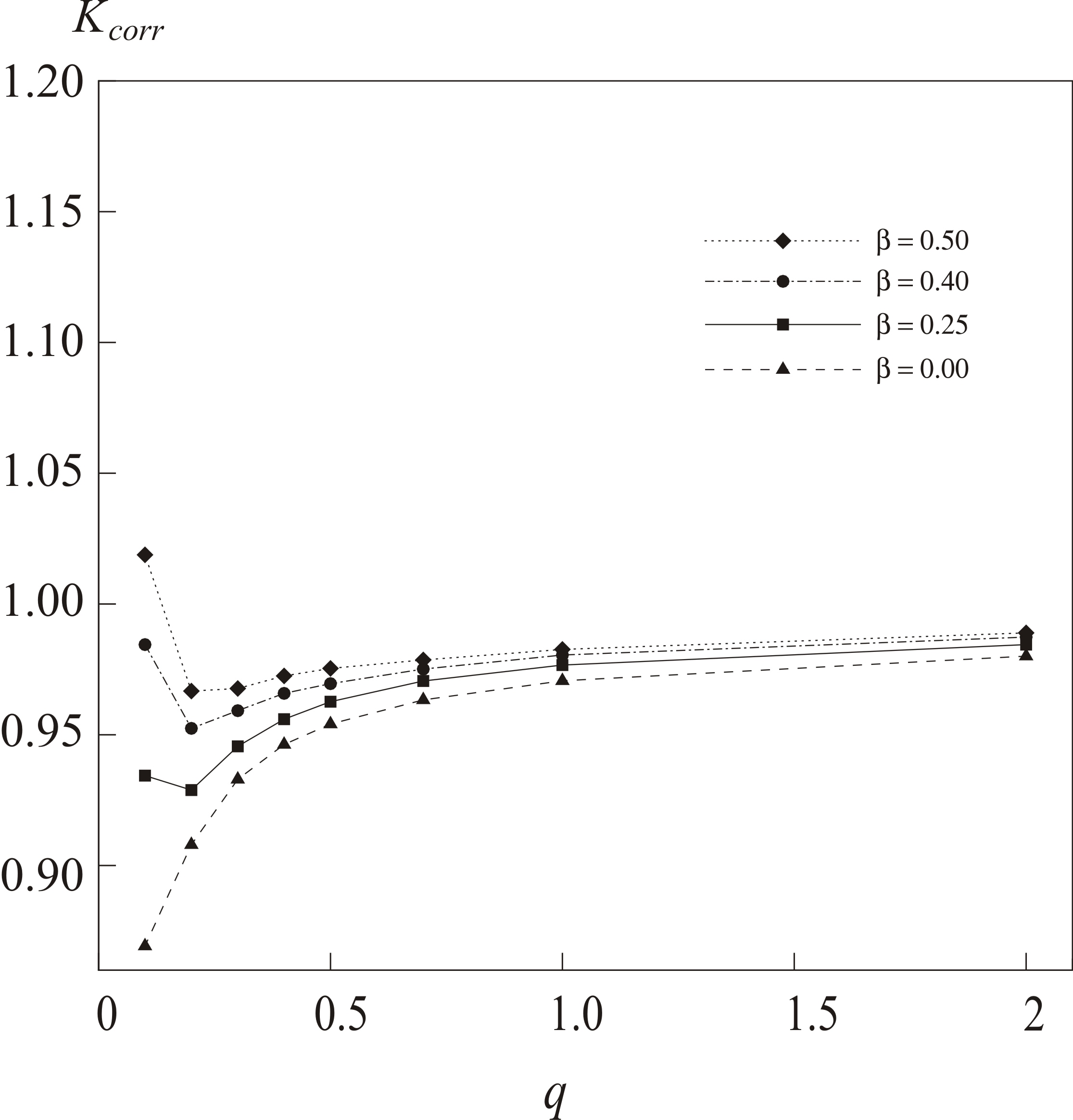}}
\caption{The K-corrections as functions of $q$ for various gravitational darkening coefficients $\beta$.  Here, $\mu=1.0$, $i=90^{o}$; the other model parameters are given in Table \ref{tabular:model}. The calculations were made by applying Algorithm I to the $H_{\gamma}$ line.}
\label{ris:Kuruz_beta}
\end{figure}

\begin{figure}[h]
\center{\includegraphics[width=1\linewidth]{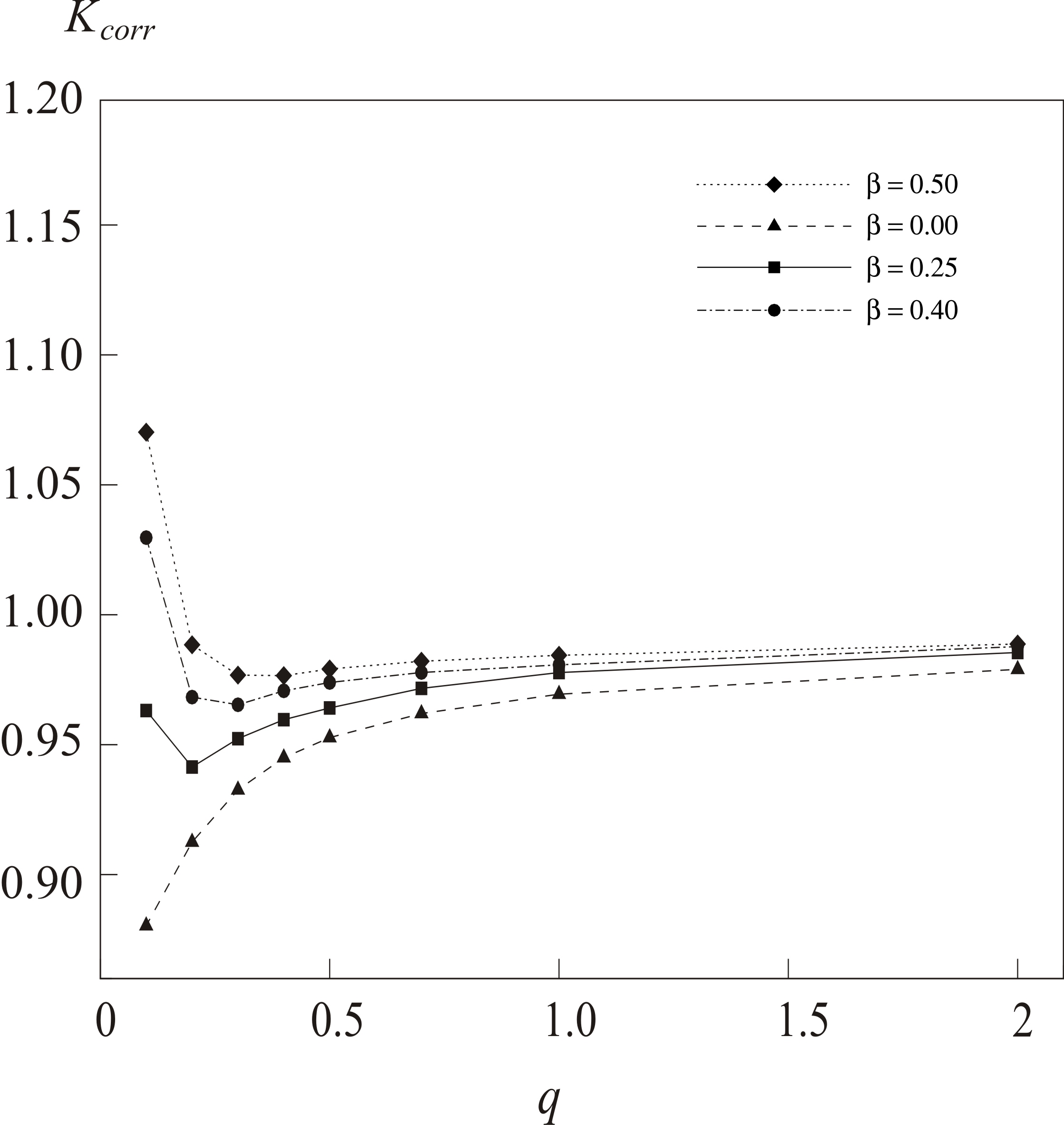}}
\caption{Same as Fig. \ref{ris:Kuruz_beta}, for the calculations made using the method of  \citet{ss5}.}
\label{ris:Wilson_beta}
\end{figure}

\begin{figure}[h]
\center{\includegraphics[width=1\linewidth]{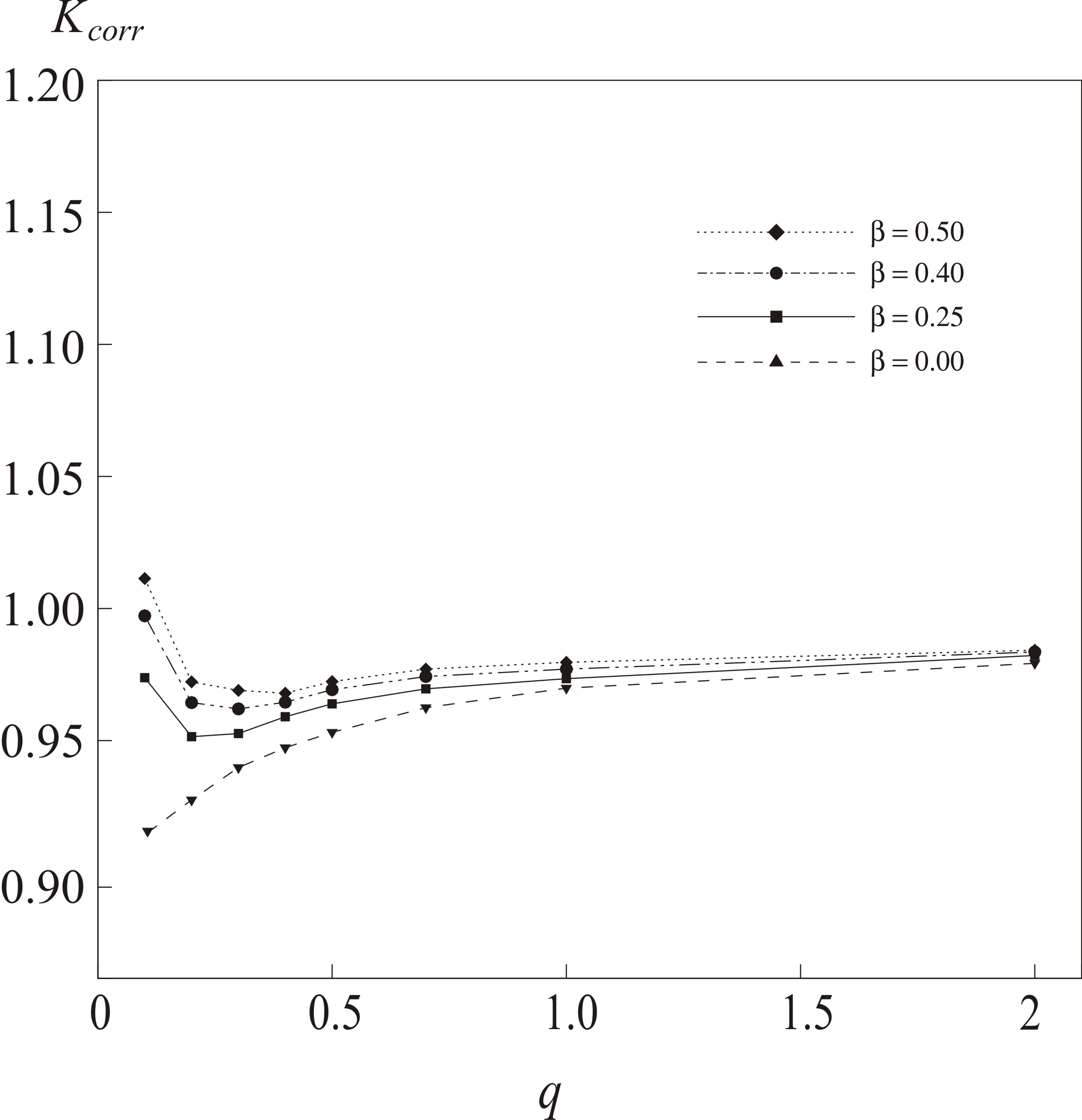}}
\caption{Same as Fig. \ref{ris:Kuruz_beta}, for calculations made by applying Algorithm II to the  $H_{\gamma}$ line. }
\label{ris:Syntes_beta}
\end{figure}

\begin{table}[H]
\caption{Parameters of the model}
\label{tabular:model}
\begin{center}
\begin{tabular}{lcp{9cm}}
\hline
\hline
Parameter  & Value   &   Definition\\[1.5ex]
\hline
%{\small  }& {\small }&{\small } \\
{\small $M_{x}$, $M_{\odot}$ }& {\small 0.6 - 20}&  { \small Mass of the relativistic component }\\[1.3ex]
{\small $M_{v}$, $M_{\odot}$}& {\small 10}&  { \small Mass of the optical star}\\[1.3ex]
{\small $q$}&{\small 0.06 - 2}&{\small Component mass ratio}\\[1.3ex]
{\small $\mu$}&{\small 0.85 - 1.0}&{\small Roche lobe filling factor}\\[1.3ex]
{\small $T_{eff}$, K }&{\small 22000}&{\small Effective temperature of the star} \\[1.3ex]
{\small $i$ (deg.) }&{\small 30 - 90}&{\small Orbital inclination}\\[1.3ex]
{\small $k_{x} = L_{x}/L_{v}$}&{\small 0}&{\small Ratio of the X-ray luminosity of the relativistic component to the bolometric luminosity of the star}\\[1.3ex]
{\small $\beta$}&{\small 0 - 0.5}&{\small Gravitational-darkening coefficient}\\[1.3ex]
{\small $A$ }& {0.5 } & {\small X-ray reprocessing coefficient} \\[1.3ex]
{\small $u$} & {\small 0.3} &{\small Limb-darkening coefficient (linear law)} \\[1.3ex]
{\small $e$} &{\small 0}& {\small Orbital eccentricity }\\[1.3ex]
{\small $\omega$ (deg.) }&{\small 0} &{\small Longitude of periastron of the star} \\[1.5ex]
{\small $V_{\gamma}$, km/s }&{\small 0}&{\small Radial velocity of the binary barycenter}\\[1.3ex]
{\small $P$, days }& {\small 1}&{\small Period} \\[1.3ex]
\hline
\hline
\end{tabular}
\end{center}
\end{table}
Three methods for calculating the radial velocity curves were used in our modelling: Algorithm I of \citet{ss11, ss16}, Algorithm II
of  \citet{ss9}, and the algorithm of \citet{ss5}. Figures  \ref{ris:Kuruz_beta}-\ref{ris:Syntes_beta} show the computed results for
the K-corrections as a function of $q$ obtained using the three techniques for the radial velocity calculations. The calculations were carried for $\beta$ =0, 0.25, 0.4 and 0.5. The star was assumed to fill its Roche lobe ($\mu=1$), and the orbital inclination was taken to be $90^{o}$; the other parameters are given in Table \ref{tabular:model}.  Since the different methods yielded similar results, we carried out our further calculations using Algorithm I, together with the theoretical profiles of Balmer lines tabulated
by  \citet{ss22}. The small differences in the graphs in Figs.  \ref{ris:Kuruz_beta}-\ref{ris:Syntes_beta} are due to calculational uncertainties
arising when the radial velocities are determined for very low component-mass ratios, $q\sim0.05-0.1$. For these low $q$, the radial velocity semiamplitudes are
only $\sim$ 20 - 25 km/s.

From the figures it is evident that the K-corrections depend significantly on $q$, and the K-correction–-$q$ graph has two characteristic regions. 

When $q > 0.2$, the K-correction is lower than unity and decreases monotonously with decreasing $q$, achieving a minimum when $q = q_{crit}$. In this case,
the radial velocity semiamplitude of the star in the Roche model is lower than that of the stellar barycenter ($K_{v}^{Roche} < K_{v}^{c} $). 

When $q\sim0.05-0.2$ the K-correction increases and can become greater than unity, e.g., the radial velocity semiamplitude of the star in the Roche model begins to increase for small $q$, and can become greater than the radial velocity semiamplitude of the
stellar barycenter ($K_{v}^{Roche} > K_{v}^{point} $). Note that, when $q < 1$, the binary barycenter is inside the optical star. This may be important when we consider binaries
with strong X-ray heating. In this case, the radial velocity curves of the star and of the barycenter can be appreciably different. This behavior of the K-corrections can be explained by changes in the relative locations of the binary barycenter, the stellar
barycenter, and the brightest regions of formation of the $H_{\gamma}$ line on the tidally deformed optical star. 

\begin{figure}[h]
\center{\includegraphics[width=1\linewidth]{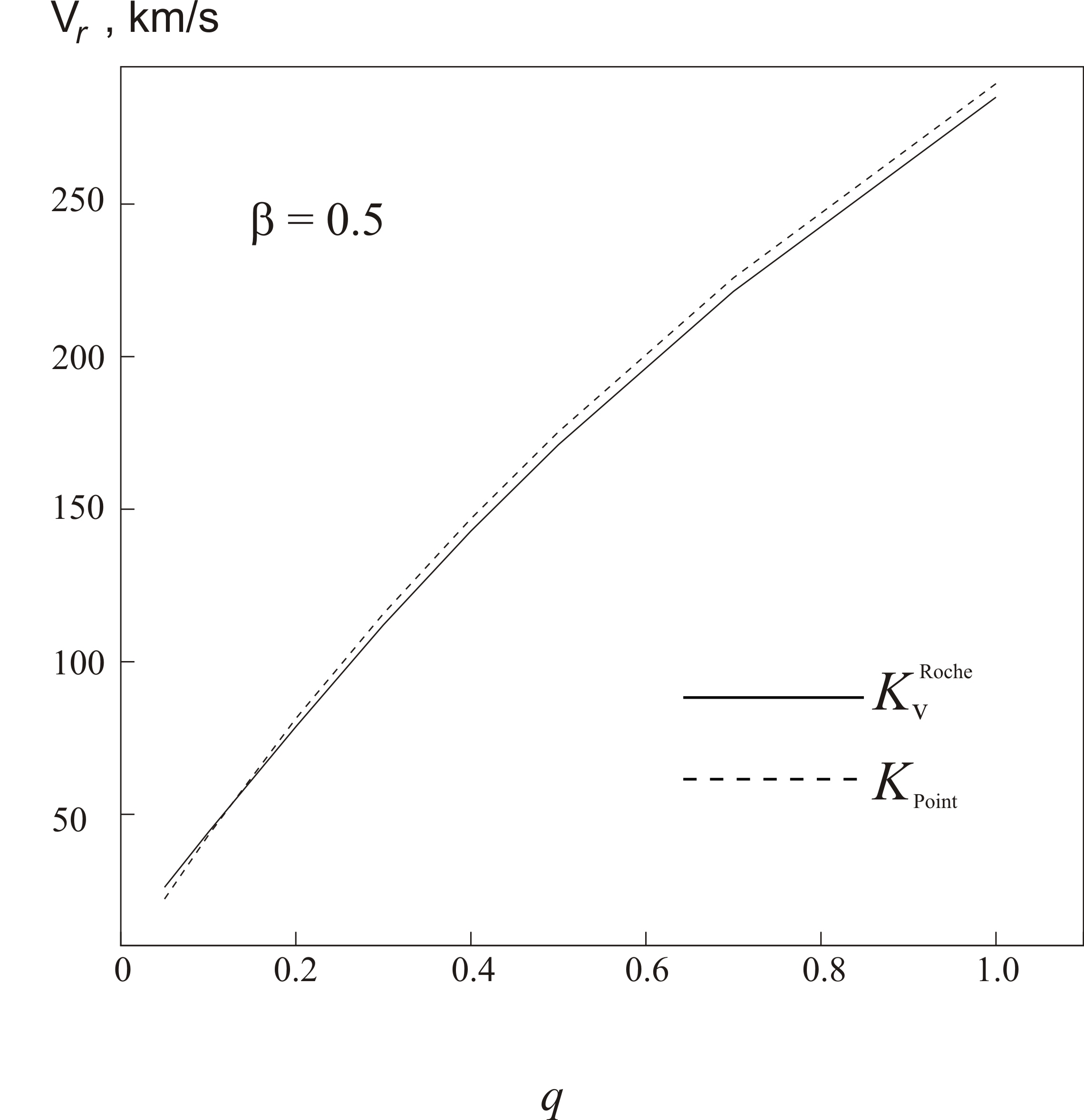}}
\caption{The radial velocity semiamplitudes as functions of $q$. The solid curve is the semiamplitude of the stellar radial velocity
curve in the Roche model, and the dashed curve the radial velocity semiamplitude of the system barycenter. Here, $\beta = 0.5$,$\mu = 1.0$, and $i = 90^{o}$; the other model parameters are given in Table \ref{tabular:model}. The calculations were made by applying Algorithm I to
the $H_{\gamma}$ line.}
\label{ris:Vr_beta05}
\end{figure}

If the temperature distribution is highly inhomogeneous ($\beta=0.5$) the contribution of the cooler regions at the “nose” of the star is small, and the radial velocity
semiamplitude for the Roche model is close to that for the point-mass model (Fig. \ref{ris:Vr_beta05}). 

\begin{figure}[h]
\center{\includegraphics[width=1\linewidth]{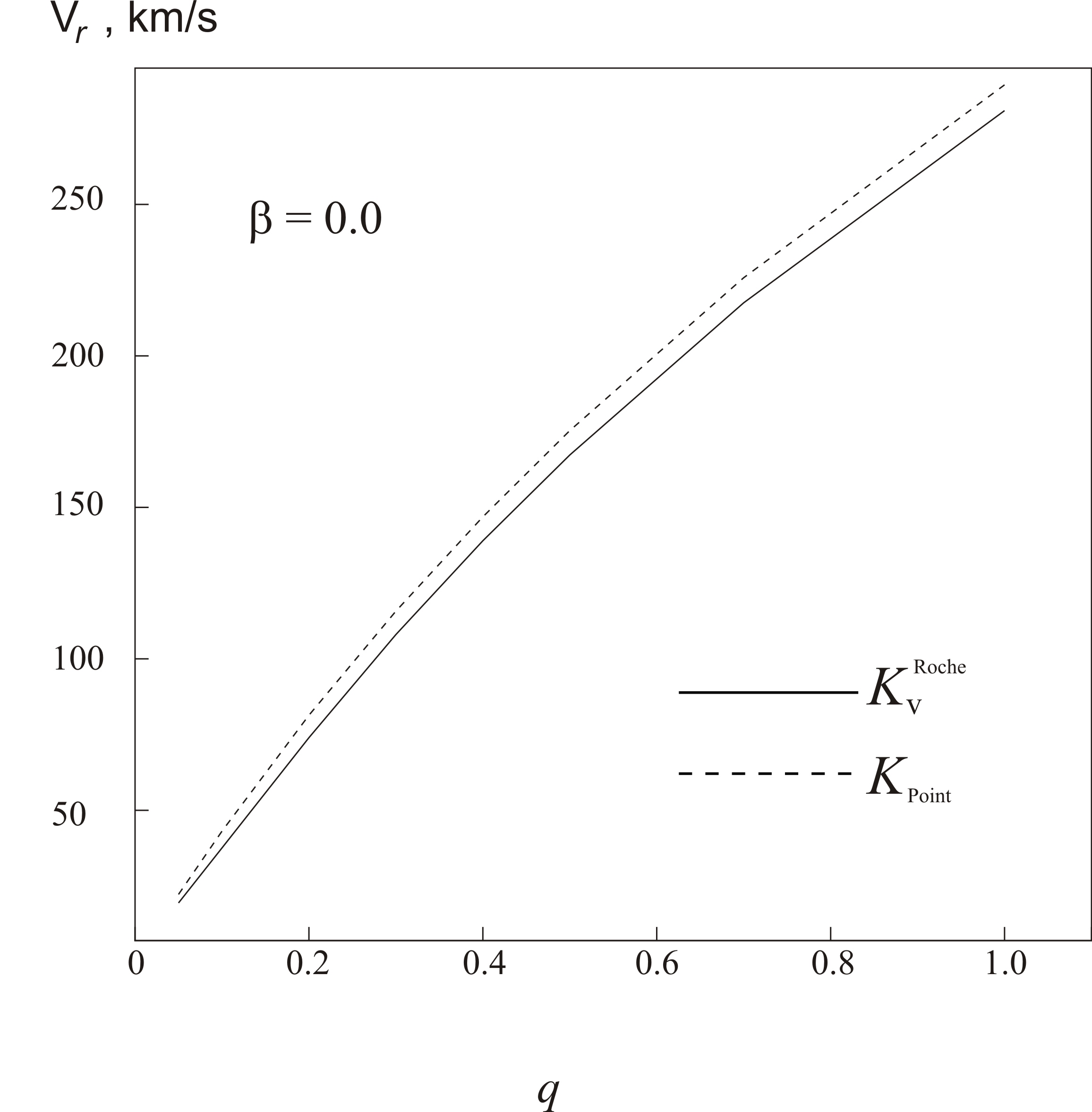}}
\caption{Same as Fig. \ref{ris:Vr_beta05} for $\beta = 0$.}
\label{ris:Vr_beta0}
\end{figure}

When the temperature is distributed homogeneously over the stellar surface ($\beta=0$) the regions
that are close to the inner Lagrangian point begin to contribute more to the total stellar radial velocity. These regions are closer to the binary barycenter than the stellar barycenter is, implying that the radial velocity semiamplitude in the Roche model decreases (Fig. \ref{ris:Vr_beta0}) compared to the case $\beta=0.5$.
 
Let us consider Figs. \ref{ris:Kuruz_beta}-\ref{ris:Syntes_beta}. Note that the K-corrections increase for $\beta=0.25,0.5$ (i.e., $\beta>0$) and can exceed unity in the region of small $q$ ($q\sim0.05-0.2$).

\begin{figure}[h]
\center{\includegraphics[width=1\linewidth]{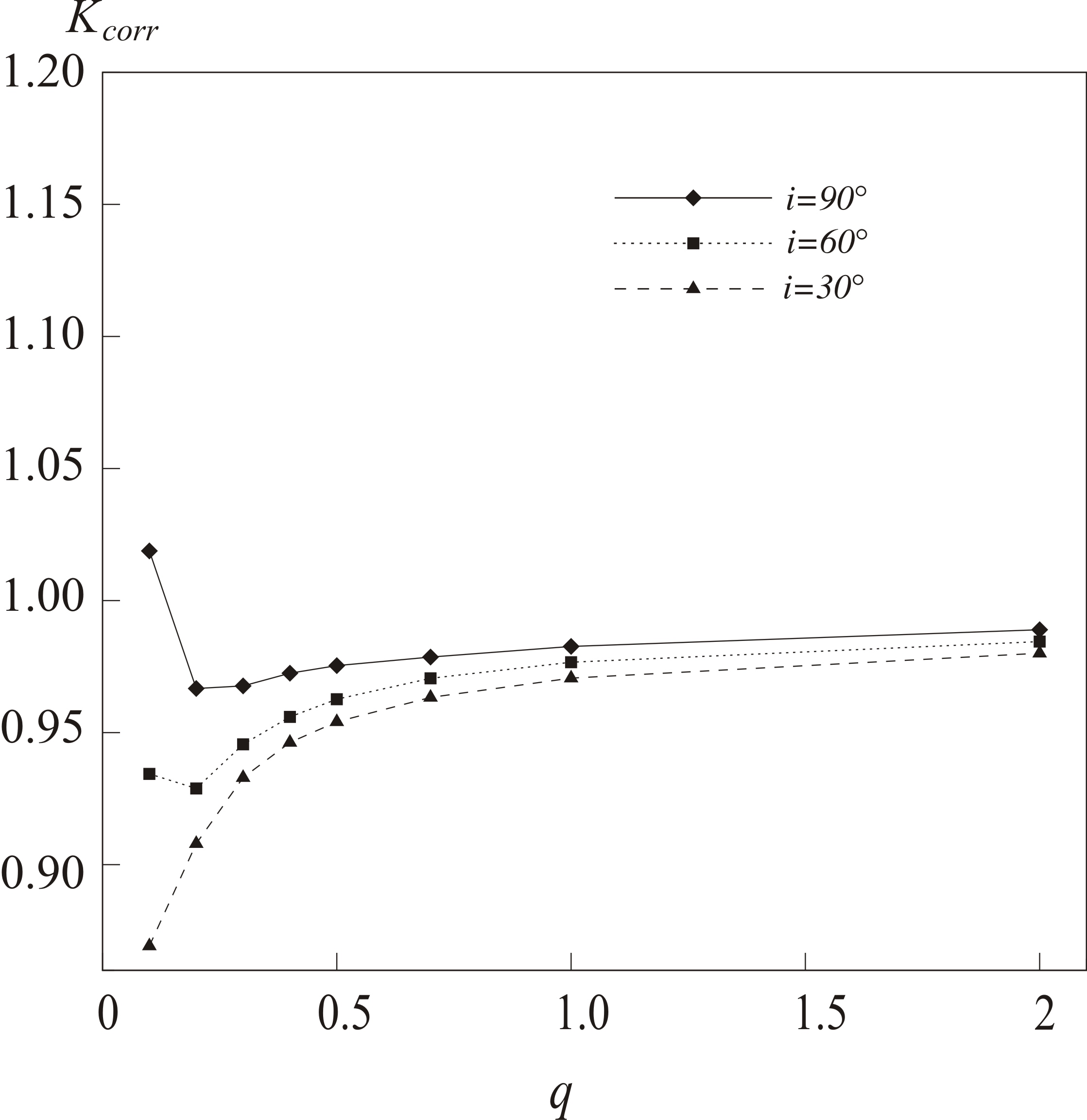}}
\caption{The K-corrections as functions of $q$ for various orbital inclinations $i$. The gravitational darkening coefficient is $\beta=0.25$,
i.e., the temperature distribution over the stellar disk is strongly inhomogeneous. Here, $\mu=1.0$; the other model parameters are given in Table \ref{tabular:model}. The calculations were made by applying Algorithm I to the $H_{\gamma}$ line.}
\label{ris:beta1_90-30}
\end{figure}

For these parameters, the maximum stellar radial velocities correspond to orbital phase $\sim0.35$, when the star is turned relative to the Earth such that its cooler region near the inner Lagrangian point is facing the observer. Hotter areas, whose velocities
are close to that of the stellar barycenter (or even higher), contribute mainly to the radiation flux at this phase. Hence, the resulting stellar radial velocities
are higher than the velocity of the stellar barycenter, $K_{v}^{Roche} > K_{v}^{point} $. 
This effect becomes more important with increasing $\beta$, since the temperature distribution over the stellar surface becomes more inhomogeneous. Moreover, the effect should increase with increasing orbital inclination $i$, which can be seen
fairly clearly in Fig. \ref{ris:beta1_90-30}. If the temperature is homogeneously
distributed over the surface ($\beta = 0$), there is no dependence on the orbital inclination, and the K-corrections approach unity with increasing $q$.

\begin{figure}[h]
\center{\includegraphics[width=1\linewidth]{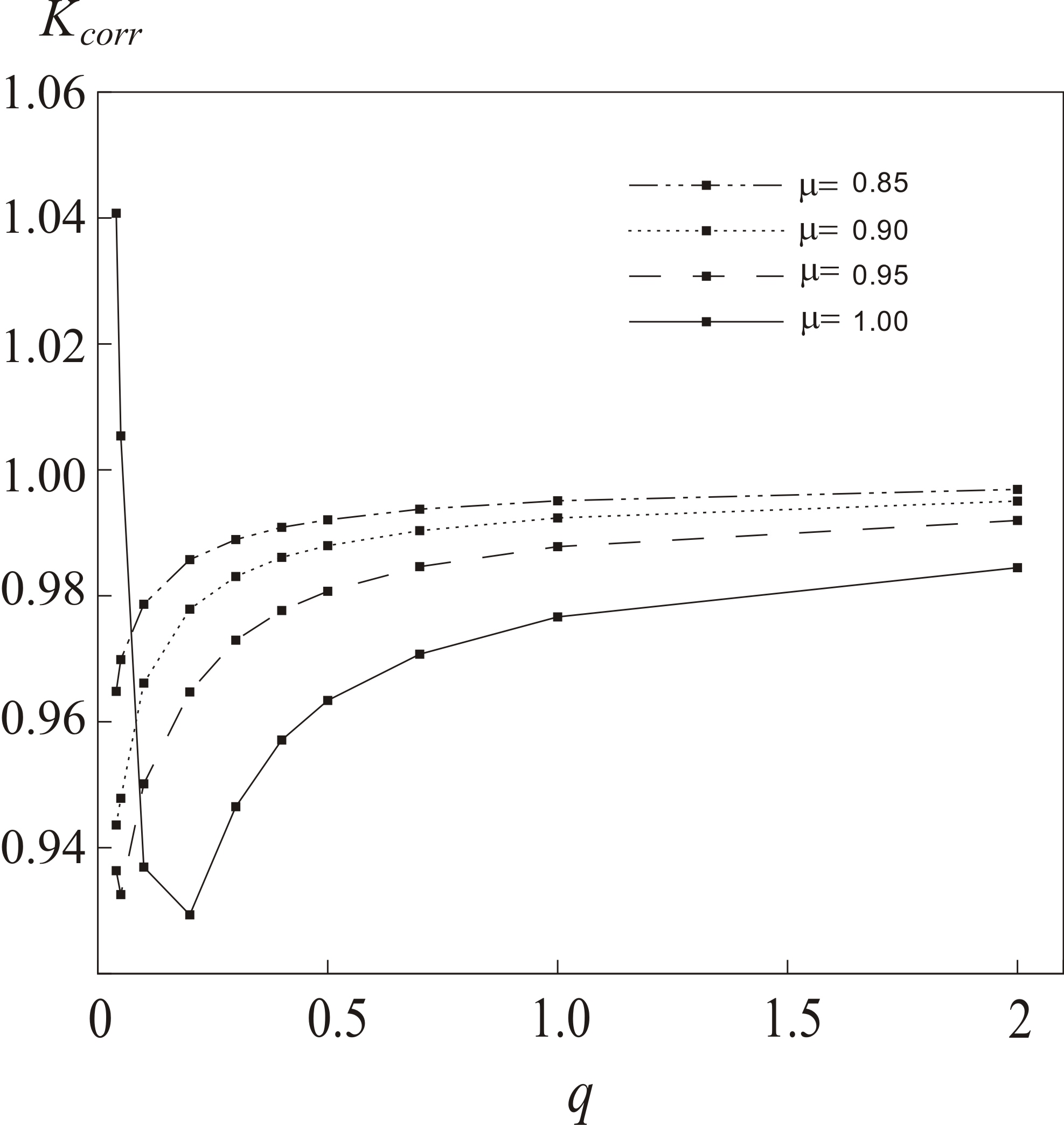}}
\caption{The K-corrections as functions of $q$ for various Roche lobe filling factors $\mu$. Here, $\beta=0.25$ and $i=90^{o}$; the other model parameters are given in Table \ref{tabular:model}. The calculations were made by applying Algorithm I to the $H_{\gamma}$ line.}
\label{ris:mu}
\end{figure}

\begin{figure}[h]
\center{\includegraphics[width=1\linewidth]{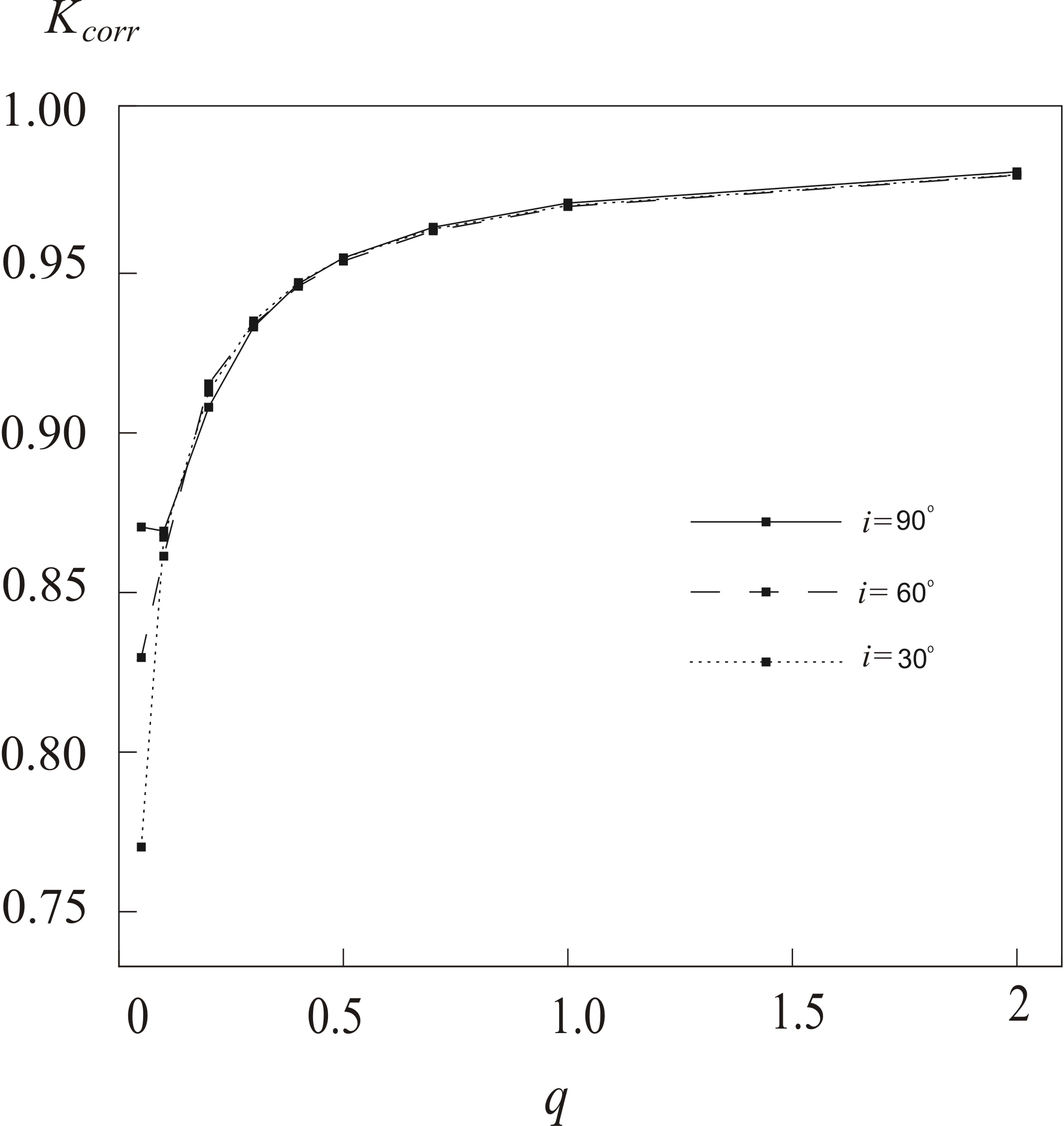}}
\caption{Same as Fig. \ref{ris:beta1_90-30} for $\beta=0.0$ (the temperature of the star is the same over the entire surface).}
\label{ris:beta0_90-30}
\end{figure}

The K-correction–-$q$ graphs for various orbital inclinations $i$ and the Roche lobe filling factors $\mu$ are presented in Figs. \ref{ris:beta1_90-30}--\ref{ris:mu}. Figures 
\ref{ris:beta1_90-30} and \ref{ris:beta0_90-30} show the plots for orbital inclinations in the range $i=30^\circ - 90 ^\circ$, for $\beta=0.5$ (Fig. \ref{ris:beta1_90-30}) and $\beta=0$ (Fig. \ref{ris:beta0_90-30}).

Figure \ref{ris:mu} shows the K-corrections as functions
of $q$ for the Roche lobe filling factors in the range $\mu=0.85 - 1.0$. The K-corrections vary especially strongly with $q$ when the star fills its Roche lobe completely
($\mu = 1.0$). Therefore, we conclude that $\mu$ can strongly
affect the K-corrections. Note that all the optical components in the analyzed binaries fill (or almost fill) their Roche lobes.

\begin{figure}[h]
\center{\includegraphics[width=1\linewidth]{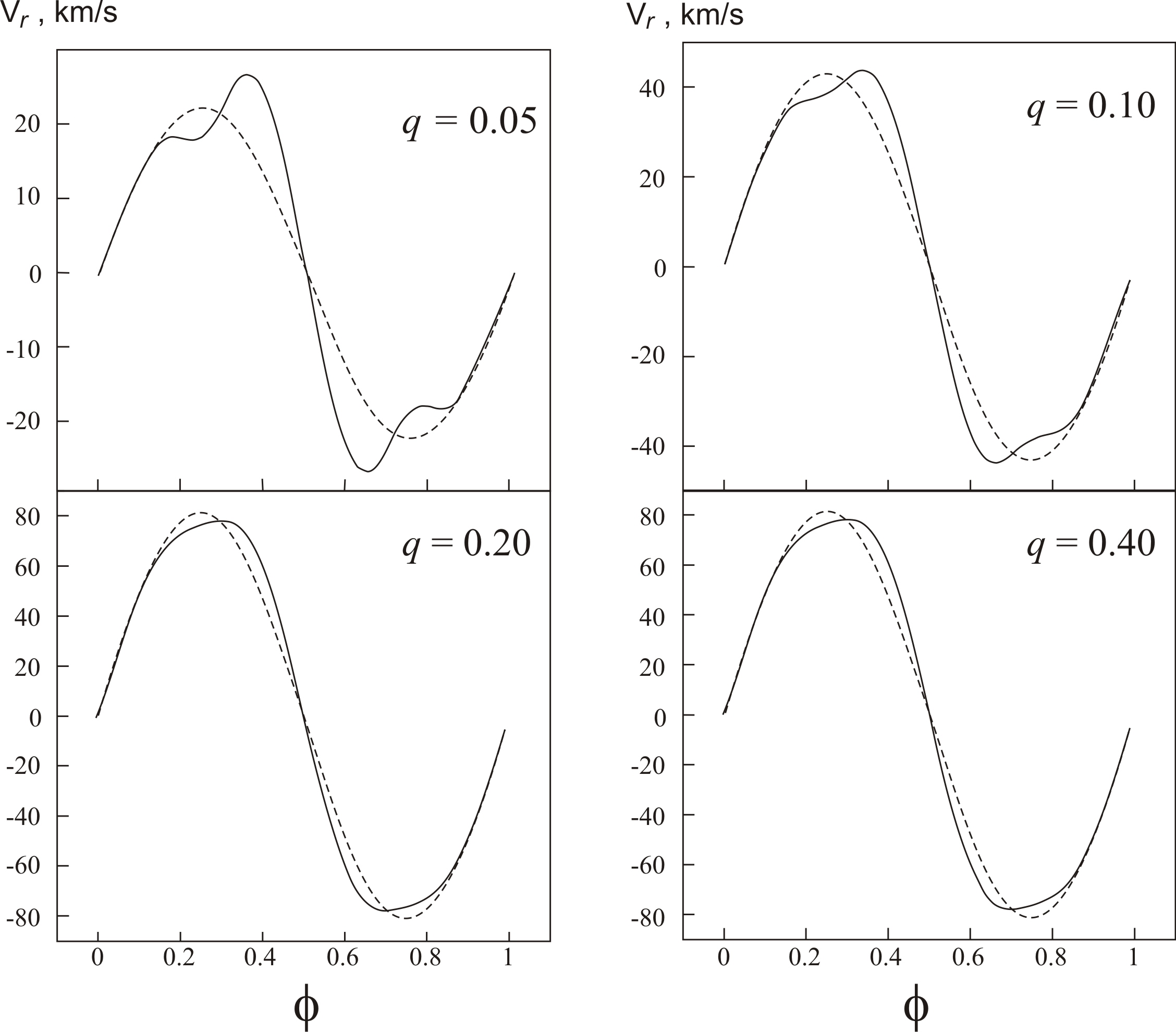}}
\caption{Shapes of the radial velocity curves as functions of the component mass ratio q. The radial velocity shapes are different
for $q < q_{crit}$ ($q = 0.05, 0.1$) and for $q\geqq_{crit} (q=0.2,0.4)$. The solid curves are the stellar radial velocity curves in the Roche model, and the dashed curves the radial velocity curves of the system barycenter. Here, $\beta=0.25$, $\mu=1.0$ and $i=90^{o}$; the other model parameters are given in Table \ref{tabular:model}. The calculations were made by applying Algorithm I to the $H_{\gamma}$ line.}
\label{ris:model3}
\end{figure}

Figure \ref{ris:model3}  shows the radial velocity curves in the Roche model (solid curves) and the radial velocity curves of the stellar barycenter (dashed curves) for various component-mass ratios in the range $q=0.05-0.4$.

Our calculations have shown that the radial velocity semiamplitudes for the star in the Roche model and for the stellar barycenter can differ appreciably. This difference affects the determination of the mass of the relativistic component. The fact that the
K-correction reaches a minimum at  $q = q_{crit}$ enables us to estimate the maximum upper uncertainty in the mass in the point-mass model.  Table \ref{tabular:q-crit} provides
the values of $q_{crit}$, and the corresponding values of $K_{corr}$ and $\Delta M_{x}/M_{x}$ for various orbital inclinations $i$.The calculations were made with fixed values of $\beta=0.25$ and $\mu=1$; the other model parameters
are given in Table \ref{tabular:model}. 
The relative underestimation of the mass of the relativistic component $\Delta M_{x}/M_{x}$ makes it clear to which extent the mass $M_x$ could be
underestimated.  Table \ref{tabular:q-crit} shows that the mass of the
relativistic component in the point-mass model can
be underestimated by $\sim30\%$. 
\begin{table}[H]
\caption{Maximum relative underestimation of the mass
$M_x$ for various orbital inclinations}
\label{tabular:q-crit}
\begin{center}
\begin{tabular}{cccc}
\hline
\hline
$i$, (deg)& $q_{crit}$&$K_{corr}$&$\Delta M_{x}/M_{x}$ \\
\hline
30&0.06&0.847&0.39\\
60&0.09&0.895&0.28\\
90&0.1&0.937&0.17\\
\hline
\hline
\end{tabular}
\end{center}
\end{table}

\section{K-CORRECTIONS FOR RADIAL VELOCITY CURVES OF MASSIVE X-RAY BINARIES WITH OB SUPERGIANTS}

\citet{ss10} studied the five eclipsing X-ray binaries Cen X-3, LMC X-4, SMC X-1, Vela X-1, and 4U 1538-52 to determine the dynamical masses of their neutron stars. They analyzed both new light curves and light curves published earlier using a
program for synthesizing light curves in the Roche model (\citealt{ss31}). The Monte Carlo method was used to determine the values of $q$, $i$, $e$ and $\omega$ for which the
durations of the X-ray eclipses were closest to the observed durations for each system. The projected semi-major axes of the neutron-star orbits $a_x \sin{i}$ derived using X-ray pulsar timing were fixed. The use of the Roche model made it possible to model
durations of the X-ray eclipses with high accuracy. The radial velocity semiamplitudes calculated in a point mass model were used as one of the fitting criteria.
These were also compared with the observed values. 

However, the observed radial velocity semiamplitude for the optical star $K_{v}$ was assumed to be fixed in \citet{ss10} when varying the parameters $q$, $\mu$ and $i$. Our calculations indicate (see Section 2) that the observed value $K_{v}$ should be regularly corrected when $q$, $\mu$ and $i$ are varied in a Roche model with fixed $K_{point}$. Or, if a fixed value of $K_{v}$ is used, the theoretical value of  $K_{point}$ should be corrected. Recall that  $K_{point}$ is the stellar radial velocity semiamplitude
calculated in the point-mass model.

\begin{figure}[h]
\center{\includegraphics[scale=0.7]{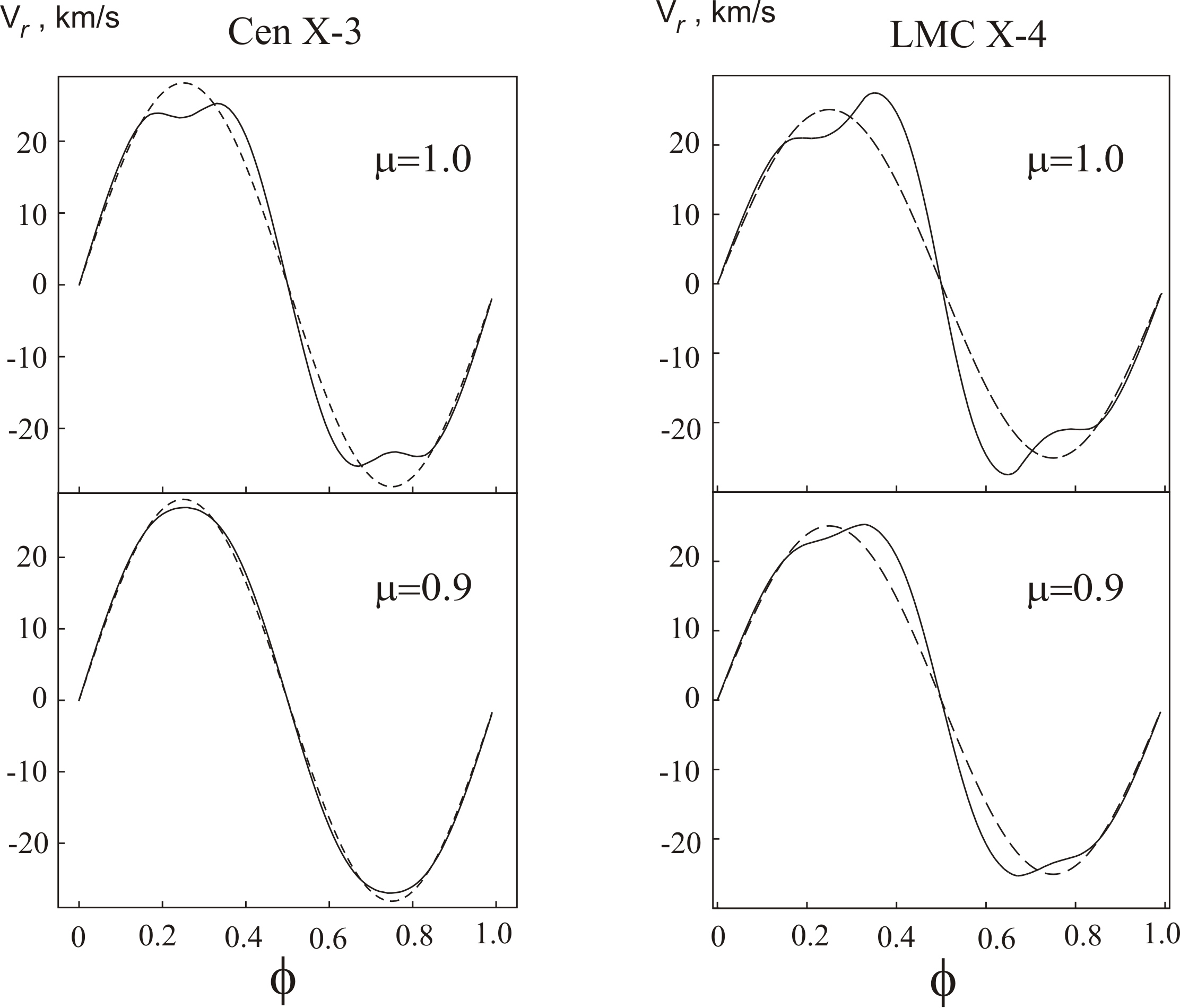}}
\caption{Model radial velocity curves for the optical stars in the binaries Cen X-3 (left) and LMC X-4 (right). The solid curves
show the stellar radial velocity curves in the Roche model and the dashed curves the radial velocity curves of the stellar the barycenter. The star fills its Roche lobe for $\mu =1$ and almost fills it for $\mu =0.9$. It was assumed that $q=0.06$ and $i=67^{o}$ for Cen X-3, and $q=0.06$ and $i=67^{o}$ for LMC X-4; the other model parameters are given in Table \ref{tabular:all}. The calculations were
carried out by applying Algorithm I to the $H_{\gamma}$ line.} 
\label{ris:CenX3} %% метка рисунка для ссылки на него
\end{figure}

\begin{figure}[h]
\center{\includegraphics[scale=0.7]{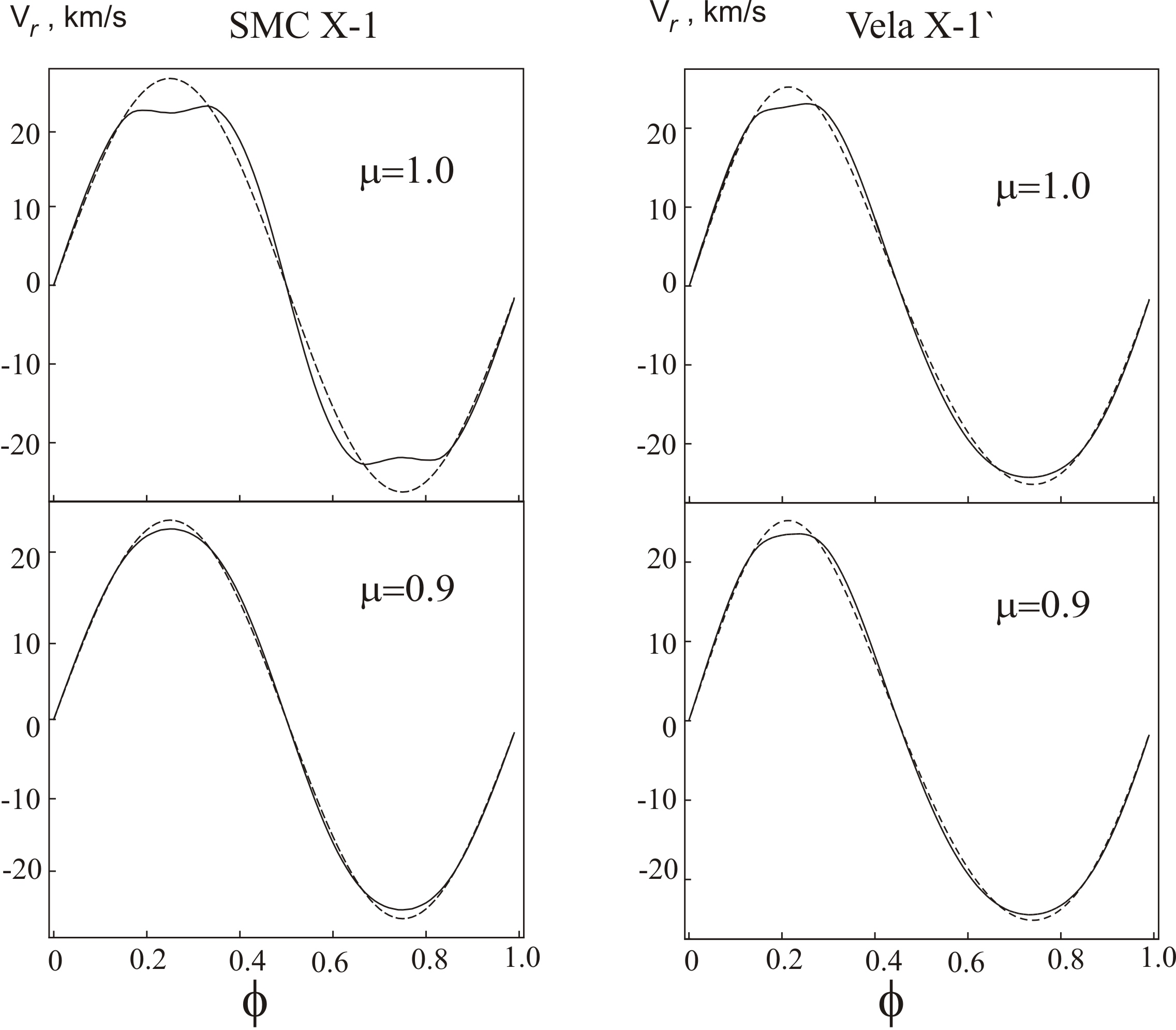}}
\caption{Same as in Fig. \ref{ris:CenX3}, for SMC X-1 (left) and Vela X-1 (right). It was assumed that $q=0.07$ and $i=68^{o}$ for SMC X-1, and $q=0.073$ and $i=77^{o}$ for Vela X-1. } 
\label{ris:VellaX1} 
\end{figure}

\begin{figure}[h]
\center{\includegraphics[scale=0.7]{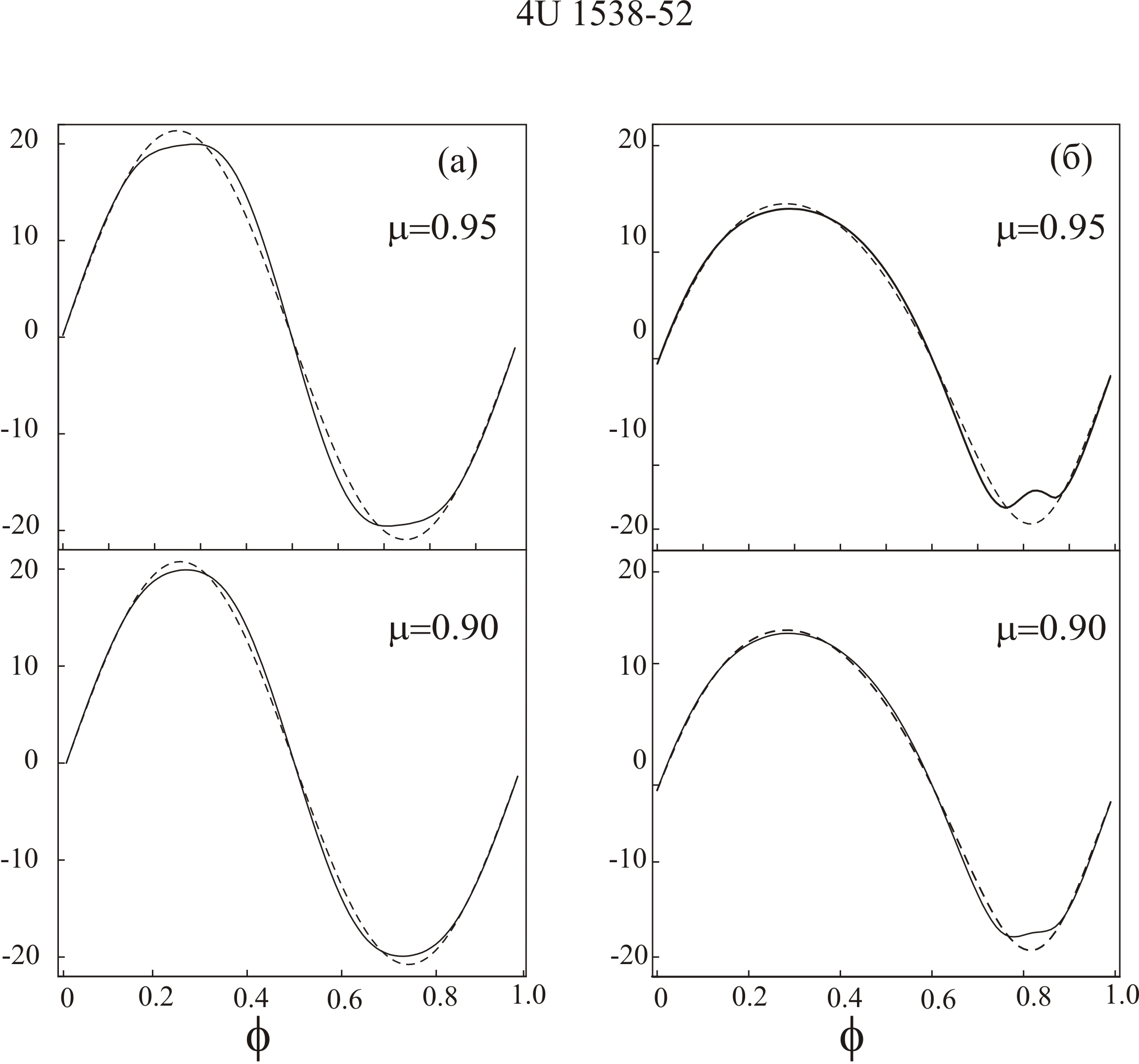}}
\caption{Same as in Fig. \ref{ris:CenX3}, for 4U 1538-52: (a) circular orbit, (b) elliptical orbit. The radial velocity curves are shown for various Roche lobe filling factors ($\mu =0.9,0.95$); it was assumed that $q=0.066$, $i=76^{o}$.} %% подпись к рисунку
\label{ris:4U_Mu095e} %% метка рисунка для ссылки на него
\end{figure}

We calculated the K-corrections for the probable ranges of $q$, $\mu$ and $i$ for each of the X-ray binaries Cen X-3, LMC X-4, SMC X-1, Vela X-1, and 4U 1538-52. The stellar radial velocity curves for the analyzed systems were modelled using the parameters
of \citet{ss10} given in Table \ref{tabular:all}.The K-corrections were
calculated using a technique that is similar to that applied for the model problem (see Section 2). The results of the radial velocity calculations are shown
in Figs \ref{ris:CenX3} - \ref{ris:4U_Mu095e}. For each X-ray binary, radial velocity
curves are provided for the stellar barycenter and the star in the Roche model. The calculations were made for two different Roche-lobe filling factors: the star
fills its Roche lobe completely if $\mu = 1$ and almost fills its Roche lobe if $\mu = 0.9$. The figures show that the maximum deviations between the stellar radial velocity
curve in the Roche model and the curve for the barycenter are obtained for $\mu = 1$, which coincides with the modeling results.

Tables of K-corrections based on numerical modelling of the radial velocity curves for the OB supergiants in the X-ray binaries Cen X-3, LMC X-4, SMC X-1, Vela X-1, and 4U 1538-52 were compiled (see Appendix, Tables \ref{tabular:KCorrectionCenX3} - \ref{tabular:KCorrection4Uc}). These correction coefficients can be used to recalculate the observed stellar radial velocity semiamplitudes to the radial velocity semiamplitudes for the barycenter.

\begin{table}[H]
\caption{Parameters for the radial velocity synthesis in the Roche model for the stars in five X-ray binaries using data available in \citet{ss10}}
\label{tabular:all}
\begin{center}
\begin{tabular}{lccccc}
\hline
\hline
&&&&&\\
Parameter & Cen X-3 & Vela X-1 & SMC X-1 & LMC X-4 & 4U1538-52 \\
&&&&&\\
\hline
$M_{x}$, $M_{\odot}$ & 1.1 - 3.3  & 1.65 - 2.1 &  0.9 - 1.35  &  0.9 - 2.4  & 0.88-1.1 \\[1.3ex]
$M_{v}$, $M_{\odot}$  & 22 & 24 & 15 & 15 & 22 \\[1.3ex]
$q$ & 0.05 - 0.15  & 0.069 - 0.087 &  0.06 - 0.09  &  0.06 - 0.16  & 0.044 - 0.05 \\[1.3ex]
$\mu$ & 0.9-1.0 & 0.95-1.0 & 0.9-1.0 & 0.9-1.0 & 0.9-1.0 \\[1.3ex]
$T_{eff}$, K & 30000 & 25000 & 25000 & 30000 & 30000 \\[1.3ex]
$i$, deg & 65-80 & 77-83 & 65-70 & 65-70 & 72-80 \\[1.3ex]
$k_{x}$& 0.05 & 0.003 & 0.25 & 1.4 & 0.05 \\[1.3ex]
$\beta$ & 0.25 & 0.25 & 0.25 & 0.25 & 0.25\\[1.3ex]
$A$ & 0.5 & 0.5 & 0.5 & 0.5 & 0.5 \\[1.3ex]
$u$ & 0.3 & 0.3 & 0.3 & 0.3 & 0.3 \\[1.3ex]
$e$ & 0 & 0.0898 & 0 & 0 & 0 \\[1.3ex]
$\omega$, deg & 0 & 332.59 & 0 & 0 & 0\\[1.3ex]
$V_{\gamma}$, km/s & 0 & 0 & 0 & 0 & 0\\[1.3ex]
$P$, days & 2.087 & 8.964 & 3.892 & 1.408 & 2.087\\[1.3ex]
\hline
\hline
\end{tabular}
\end{center}
\end{table}

\section{CONCLUSIONS}
We have calculated K-corrections for the radial velocity curve of a star in a Roche model as functions of the parameters $q$, $\beta$, $\mu$, $i$. We have shown
that the existence of a minimum in the K-corrections at $q = q_{crit}$ enables estimation of the maximum upper uncertainty of the mass in a point-mass model.

We have presented tabulated K-corrections for probable ranges of $q$, $\mu$, $i$ for the X-ray binaries Cen X-3, LMC X-4, SMC X-1, Vela X-1, and 4U 1538-52. These tables may be helpful in more correct determinations of the masses of the X-ray binary components using the Monte Carlo method. The tables indicate that the masses of the neutron stars in the analyzed X-ray binaries determined using models with fixed radial velocity semiamplitudes
are underestimated.

The behaviour of the K-corrections for small $q$ and the presence of a region where the K-corrections exceed unity is striking. If the compact objects are neutron stars, such $q$ values are characteristic of X-ray binaries whose optical components have masses
of approximately 20~$M_{\odot}$ and higher.

We hope that the use of these tables of K-corrections will make it possible to correctly take into account the effects of the nearness the components in these systems. This should diminish the uncertainties arising when the masses of compact objects are determined using point mass models.

\section{ACKNOWLEDGMENTS} 
This work was supported by the Russian Foundation for Basic Research (project 11-02-00258) and the Program of Support to Leading Scientific Schools of the Russian Federation (NSh-2374.2012.2).
\bibliographystyle{mn2e}
\bibliography{Mainkcorr} 

\newpage
\section{Appendix}

%\begin{figure}[h]
%\center{\includegraphics[scale=0.55]{SMCX-1all.jpg}}
%\caption{Кривые лучевых скоростей оптической звезды в системе SMC X-1. Сплошная линия - кривая лучевых скоростей звезды в моделе Роша. Пунктирная линия - кривая лучевых скоростей центра масс звезды. а: Звезда заполняет свою полость Роша ($\mu =1$). б: Звезда близка к заполнению полости Роша ($\mu =0.9$). Расчеты выполнены с Алгоритмом I. Принято: $q=0.07$, $i=68^{o}$. Остальные параметры модели приведены в табл. \ref{tabular:all} .} %% подпись к рисунку
%\label{ris:SMCX1} %% метка рисунка для ссылки на него
%\end{figure}

%\begin{figure}[h]
%\center{\includegraphics[scale=0.55]{LMC_X-4_all.jpg}}
%\caption{Кривые лучевых скоростей оптической звезды в системе LMC X-4. Сплошная линия - кривая лучевых скоростей звезды в моделе Роша. Пунктирная линия - кривая лучевых скоростей центра масс звезды. а- Звезда заполняет свою полость Роша ($\mu =1$). б- Звезда близка к заполнению полости Роша ($\mu =0.95$). Расчеты выполнены с Алгоритмом I. Принято: $q=0.085$, $i=68^{o}$. Остальные параметры модели приведены в табл. \ref{tabular:all} .} %% подпись к рисунку
%\label{ris:LMCX-4Mu1_0} %% метка рисунка для ссылки на него
%\end{figure}

%\begin{figure}[h]
%\center{\includegraphics[scale=0.55]{4U_Mu_all_c.jpg}}
%\caption{Кривые лучевых скоростей оптической звезды в системе %4U1538-52 в случае круговой орбиты. Сплошная линия - кривая лучевых %скоростей звезды в моделе Роша. Пунктирная линия - кривая лучевых %скоростей центра масс звезды. а: Звезда заполняет свою полость Роша %($\mu =1$). б: Звезда близка к заполнению полости Роша ($\mu %=0.9$). Расчеты выполнены с Алгоритмом I. Принято: $q=0.048$, %$i=76^{o}$. Остальные параметры модели приведены в табл. %\ref{tabular:all} .} %% подпись к рисунку
%\label{ris:4U_Mu095c} %% метка рисунка для ссылки на него
%\end{figure}

\newpage
\section{THE TABLES OF K-CORRECTIONS}

\begin{table}[h]

\caption{K-corrections for the optical star in Cen X-3}

\label{tabular:KCorrectionCenX3}

\begin{center}

\begin{tabular}{p{40pt}p{50pt}p{50pt}p{50pt}p{50pt}}

\hline

\hline \multicolumn{5}{c}{$\mu=1.00$} \\

\hline q & \textit{i}=$60^{\circ}$& \textit{i}=$65^{\circ}$ &
\textit{i}=$70^{\circ}$ & \textit{i}=$80^{\circ}$ \\

\hline

0.050 & 0.87908 & 0.89517 & 0.91067 & 0.95020 \\

0.060 & 0.87505 & 0.88839 & 0.90103 & 0.93115 \\

0.067 & 0.87114 & 0.88680 & 0.89829 & 0.92094 \\

0.070 & 0.87085 & 0.88505 & 0.89694 & 0.91726 \\

0.090 & 0.87591 & 0.88475 & 0.89127 & 0.90289 \\

0.100 & 0.88266 & 0.88811 & 0.89258 & 0.90074 \\

0.150 & 0.91164 & 0.91247 & 0.91524 & 0.90481 \\

\hline \multicolumn{5}{c}{$\mu=0.95$} \\

\hline q & \textit{i}=$60^{\circ}$& \textit{i}=$65^{\circ}$ &
\textit{i}=$70^{\circ}$ & \textit{i}=$80^{\circ}$ \\

\hline

0.050 & 0.91786 & 0.91760 & 0.92008 & 0.91880 \\

0.060 & 0.92597 & 0.92518 & 0.92823 & 0.92551 \\

0.067 & 0.92940 & 0.92994 & 0.93279 & 0.92976 \\

0.070 & 0.93137 & 0.93124 & 0.93367 & 0.93186 \\

0.090 & 0.94101 & 0.94168 & 0.94108 & 0.94327 \\

0.100 & 0.94423 & 0.94493 & 0.94471 & 0.94702 \\

0.150 & 0.95699 & 0.95805 & 0.95787 & 0.95853 \\

\hline \multicolumn{5}{c}{$\mu=0.90$} \\

\hline q & \textit{i}=$60^{\circ}$& \textit{i}=$65^{\circ}$ &
\textit{i}=$70^{\circ}$ & \textit{i}=$80^{\circ}$ \\

\hline

0.050 & 0.95153 & 0.95222 & 0.95252 & 0.95424 \\

0.060 & 0.95593 & 0.95626 & 0.95702 & 0.95786 \\

0.067 & 0.95795 & 0.95944 & 0.95946 & 0.95928 \\

0.070 & 0.95941 & 0.96051 & 0.96020 & 0.95977 \\

0.090 & 0.96455 & 0.96557 & 0.96599 & 0.96524 \\

0.100 & 0.96659 & 0.96731 & 0.96799 & 0.96784 \\

0.150 & 0.97398 & 0.97410 & 0.97485 & 0.97521 \\

\hline

\hline

\end{tabular}

\end{center}

\end{table}

\begin{table}[h]

\caption{K-corrections for the optical star in SMC X-1 }

\label{tabular:KCorrectionSMCX1}

\begin{center}

\begin{tabular}{p{40pt}p{50pt}p{50pt}p{50pt}p{50pt}}

\hline

\hline \multicolumn{5}{c}{$\mu=1.00$} \\

\hline q & \textit{i}=$60^{\circ}$& \textit{i}=$65^{\circ}$ &
\textit{i}=$68^{\circ}$ & \textit{i}=$70^{\circ}$ \\

\hline

0.060 & 0.84560 & 0.86049 & 0.86752 & 0.87204 \\

0.070 & 0.84830 & 0.86095 & 0.86699 & 0.87019 \\

0.071 & 0.84964 & 0.86027 & 0.86673 & 0.87083 \\

0.074 & 0.85120 & 0.86015 & 0.86609 & 0.87055 \\

0.076 & 0.85210 & 0.86138 & 0.86676 & 0.87027 \\

0.080 & 0.85416 & 0.86235 & 0.86715 & 0.87060 \\

0.090 & 0.86550 & 0.86796 & 0.87059 & 0.87266 \\

\hline \multicolumn{5}{c}{$\mu=0.95$} \\

\hline q & \textit{i}=$60^{\circ}$& \textit{i}=$65^{\circ}$ &
\textit{i}=$68^{\circ}$ & \textit{i}=$70^{\circ}$ \\

\hline

0.060 & 0.91682 & 0.91824 & 0.91783 & 0.91782 \\

0.070 & 0.92466 & 0.92604 & 0.92530 & 0.92534 \\

0.071 & 0.92559 & 0.92600 & 0.92623 & 0.92673 \\

0.074 & 0.92707 & 0.92844 & 0.92916 & 0.92828 \\

0.076 & 0.92796 & 0.92978 & 0.92959 & 0.93008 \\

0.080 & 0.93049 & 0.93226 & 0.93294 & 0.93258 \\

0.090 & 0.93499 & 0.93748 & 0.93738 & 0.93820 \\

\hline \multicolumn{5}{c}{$\mu=0.90$} \\

\hline q & \textit{i}=$60^{\circ}$& \textit{i}=$65^{\circ}$ &
\textit{i}=$68^{\circ}$ & \textit{i}=$70^{\circ}$ \\

\hline

0.060 & 0.94794 & 0.95083 & 0.95081 & 0.95201 \\

0.070 & 0.95098 & 0.95414 & 0.95518 & 0.95578 \\

0.071 & 0.95107 & 0.95472 & 0.95574 & 0.95632 \\

0.074 & 0.95301 & 0.95557 & 0.95658 & 0.95670 \\

0.076 & 0.95324 & 0.95622 & 0.95722 & 0.95822 \\

0.080 & 0.95457 & 0.95788 & 0.95925 & 0.95896 \\

0.090 & 0.95733 & 0.96078 & 0.96091 & 0.96180 \\

\hline

\hline

\end{tabular}

\end{center}

\end{table}

\begin{table}[h]

\caption{K-corrections for the optical star in LMC X-4}

\label{tabular:KCorrectionLMCX4}

\begin{center}

\begin{tabular}{p{40pt}p{50pt}p{50pt}p{50pt}p{50pt}}

\hline

\hline \multicolumn{5}{c}{$\mu=1.00$} \\

\hline q & \textit{i}=$60^{\circ}$& \textit{i}=$65^{\circ}$ &
\textit{i}=$68^{\circ}$ & \textit{i}=$70^{\circ}$ \\

\hline

0.060 & 1.00384 & 1.06314 & 1.09518 & 1.11198 \\

0.080 & 0.94432 & 0.98082 & 1.00393 & 1.01850 \\

0.090 & 0.93427 & 0.96645 & 0.98515 & 0.99834 \\

0.100 & 0.91924 & 0.94162 & 0.95273 & 0.96206 \\

0.120 & 0.91066 & 0.91948 & 0.92378 & 0.92725 \\

0.140 & 0.90487 & 0.90781 & 0.90810 & 0.90879 \\

0.160 & 0.90679 & 0.90607 & 0.90390 & 0.90315 \\

\hline \multicolumn{5}{c}{$\mu=0.95$} \\

\hline q & \textit{i}=$60^{\circ}$& \textit{i}=$65^{\circ}$ &
\textit{i}=$68^{\circ}$ & \textit{i}=$70^{\circ}$ \\

\hline

0.060 & 0.96633 & 0.99715 & 1.00876 & 1.01179 \\

0.080 & 0.95047 & 0.96071 & 0.96311 & 0.96479 \\

0.090 & 0.95048 & 0.95469 & 0.95600 & 0.95686 \\

0.100 & 0.94913 & 0.94914 & 0.95028 & 0.95046 \\

0.120 & 0.94914 & 0.94865 & 0.94960 & 0.94946 \\

0.140 & 0.95148 & 0.94960 & 0.95109 & 0.95192 \\

0.160 & 0.95382 & 0.95263 & 0.95306 & 0.95368 \\

\hline \multicolumn{5}{c}{$\mu=0.90$} \\

\hline q & \textit{i}=$60^{\circ}$& \textit{i}=$65^{\circ}$ &
\textit{i}=$68^{\circ}$ & \textit{i}=$70^{\circ}$ \\

\hline

0.060 & 0.96505 & 0.96864 & 0.97133 & 0.97210 \\

0.080 & 0.96342 & 0.96287 & 0.96281 & 0.96270 \\

0.090 & 0.96459 & 0.96272 & 0.96216 & 0.96211 \\

0.100 & 0.96592 & 0.96367 & 0.96302 & 0.96278 \\

0.120 & 0.96749 & 0.96577 & 0.96530 & 0.96495 \\

0.140 & 0.96893 & 0.96774 & 0.96722 & 0.96730 \\

0.160 & 0.97046 & 0.96966 & 0.96955 & 0.96933 \\

\hline

\hline

\end{tabular}

\end{center}

\end{table}

\begin{table}[h]

\caption{K-corrections for the optical star in Vela X-1}

\label{tabular:KCorrectionVelaX1}

\begin{center}

\begin{tabular}{p{40pt}p{50pt}p{50pt}p{50pt}p{50pt}}

\hline

\hline \multicolumn{5}{c}{$\mu=1.00$} \\

\hline q & \textit{i}=$73^{\circ}$& \textit{i}=$77^{\circ}$ &
\textit{i}=$80^{\circ}$ & \textit{i}=$83^{\circ}$ \\

\hline

0.06875 & 0.91828 & 0.92027 & 0.92160 & 0.92315 \\

0.07083 & 0.91911 & 0.92064 & 0.92239 & 0.92301 \\

0.07292 & 0.91948 & 0.92188 & 0.92269 & 0.92332 \\

0.07500 & 0.92069 & 0.92260 & 0.92342 & 0.92402 \\

0.07917 & 0.92208 & 0.92393 & 0.92350 & 0.92574 \\

0.08333 & 0.92330 & 0.92554 & 0.92591 & 0.94120 \\

0.08750 & 0.92559 & 0.92658 & 0.92772 & 0.92792 \\

\hline \multicolumn{5}{c}{$\mu=0.97$} \\

\hline q & \textit{i}=$73^{\circ}$& \textit{i}=$77^{\circ}$ &
\textit{i}=$80^{\circ}$ & \textit{i}=$83^{\circ}$ \\

\hline

0.06875 & 0.93561 & 0.93632 & 0.93699 & 0.93747 \\

0.07083 & 0.93645 & 0.93670 & 0.93782 & 0.93785 \\

0.07292 & 0.93680 & 0.93842 & 0.93861 & 0.93911\\

0.07500 & 0.93801 & 0.94004 & 0.94024 & 0.93983 \\

0.07917 & 0.94069 & 0.94178 & 0.94157 & 0.94201 \\

0.08333 & 0.94227 & 0.94294 & 0.94393 & 0.94358 \\

0.08750 & 0.94409 & 0.94474 & 0.94532 & 0.94575 \\

\hline \multicolumn{5}{c}{$\mu=0.95$} \\

\hline q & \textit{i}=$73^{\circ}$& \textit{i}=$77^{\circ}$ &
\textit{i}=$80^{\circ}$ & \textit{i}=$83^{\circ}$ \\

\hline

0.06875 & 0.94552 & 0.94701 & 0.94661 & 0.94749 \\

0.07083 & 0.94656 & 0.94757 & 0.94764 & 0.94805 \\

0.07292 & 0.94757 & 0.94853 & 0.94861 & 0.94858 \\

0.07500 & 0.94850 & 0.94944 & 0.95042 & 0.94993 \\

0.07917 & 0.95065 & 0.95113 & 0.95124 & 0.95202 \\

0.08333 & 0.95216 & 0.95265 & 0.95314 & 0.95352 \\

0.08750 & 0.95433 & 0.95440 & 0.95488 & 0.95524 \\

\hline

\hline

\end{tabular}

\end{center}

\end{table}

\begin{table}[h]

\caption{K-corrections for the optical star in 4U 1538-54
(eccentric orbit) }

\label{tabular:KCorrection4Ue}

\begin{center}

\begin{tabular}{p{40pt}p{50pt}p{50pt}p{50pt}p{50pt}}

\hline

\hline \multicolumn{5}{c}{$\mu=1.00$} \\

\hline q & \textit{i}=$68^{\circ}$& \textit{i}=$72^{\circ}$ &
\textit{i}=$76^{\circ}$ & \textit{i}=$80^{\circ}$ \\

\hline

0.044 & 0.95029 & 0.95084 & 0.95114 & 0.95186 \\

0.045 & 0.95067 & 0.95120 & 0.95219 & 0.95156 \\

0.046 & 0.95103 & 0.95091 & 0.95254 & 0.95260 \\

0.047 & 0.95068 & 0.95323 & 0.95352 & 0.95356 \\

0.048 & 0.95169 & 0.95290 & 0.95383 & 0.95389 \\

0.049 & 0.95263 & 0.95383 & 0.95412 & 0.95480 \\

0.050 & 0.95356 & 0.95409 & 0.95441 & 0.95566 \\
\hline \multicolumn{5}{c}{$\mu=0.97$} \\
\hline q & \textit{i}=$68^{\circ}$& \textit{i}=$72^{\circ}$ &
\textit{i}=$76^{\circ}$ & \textit{i}=$80^{\circ}$ \\
\hline
0.044 & 0.96614 & 0.96629 & 0.96628 & 0.96610 \\
0.045 & 0.96617 & 0.96701 & 0.96700 & 0.96749 \\
0.046 & 0.96690 & 0.96705 & 0.96704 & 0.96753 \\
0.047 & 0.96757 & 0.96838 & 0.96837 & 0.96819 \\
0.048 & 0.96823 & 0.96839 & 0.96774 & 0.96822 \\
0.049 & 0.96885 & 0.96837 & 0.96900 & 0.96885 \\
0.050 & 0.96883 & 0.96960 & 0.96900 & 0.96944 \\
\hline \multicolumn{5}{c}{$\mu=0.95$} \\
\hline q & \textit{i}=$68^{\circ}$& \textit{i}=$72^{\circ}$ &
\textit{i}=$76^{\circ}$ & \textit{i}=$80^{\circ}$ \\
\hline
0.044 & 0.97911 & 0.97753 & 0.97798 & 0.97695 \\
0.045 & 0.98027 & 0.97801 & 0.97710 & 0.97877 \\
0.046 & 0.98000 & 0.97781 & 0.97891 & 0.97727 \\
0.047 & 0.98108 & 0.97826 & 0.97805 & 0.97901 \\
0.048 & 0.98147 & 0.97742 & 0.97849 & 0.97757 \\
0.049 & 0.98183 & 0.97913 & 0.97892 & 0.97923 \\
0.050 & 0.98155 & 0.97891 & 0.97933 & 0.97963 \\
\hline
\hline
\end{tabular}
\end{center}
\end{table}

\begin{table}[h]
\caption{K-corrections for the optical star in 4U 1538-54
(circular orbit)}
\label{tabular:KCorrection4Uc}
\begin{center}
\begin{tabular}{p{40pt}p{50pt}p{50pt}p{50pt}p{50pt}}
\hline
\hline \multicolumn{5}{c}{$\mu=1.00$} \\
\hline q & \textit{i}=$68^{\circ}$& \textit{i}=$72^{\circ}$ &
\textit{i}=$76^{\circ}$ & \textit{i}=$80^{\circ}$ \\
\hline
0.063 & 0.89440 & 0.90335 & 0.91399 & 0.92642 \\
0.064 & 0.89410 & 0.90297 & 0.91278 & 0.92433 \\
0.066 & 0.89387 & 0.90224 & 0.91045 & 0.92173 \\
0.067 & 0.89288 & 0.90119 & 0.91000 & 0.92014 \\
0.069 & 0.89238 & 0.90051 & 0.90851 & 0.91839 \\
0.070 & 0.89211 & 0.90054 & 0.90777 & 0.91726 \\
0.071 & 0.89153 & 0.89957 & 0.90773 & 0.91581 \\
\hline \multicolumn{5}{c}{$\mu=0.95$} \\
\hline q & \textit{i}=$80^{\circ}$& \textit{i}=$76^{\circ}$ & \textit{i}=$72^{\circ}$ & \textit{i}=$68^{\circ}$ \\
\hline
0.063 & 0.92833 & 0.92974 & 0.92930 & 0.92714 \\
0.064 & 0.92865 & 0.93079 & 0.93001 & 0.92786 \\
0.066 & 0.92998 & 0.93210 & 0.93101 & 0.92928 \\
0.067 & 0.93027 & 0.93238 & 0.93198 & 0.93063 \\
0.069 & 0.93047 & 0.93424 & 0.93289 & 0.93189 \\
0.070 & 0.93175 & 0.93448 & 0.93347 & 0.93186 \\
0.071 & 0.93200 & 0.93537 & 0.93405 & 0.93342 \\
\hline \multicolumn{5}{c}{$\mu=0.90$} \\
\hline q & \textit{i}=$80^{\circ}$& \textit{i}=$76^{\circ}$ & \textit{i}=$72^{\circ}$ & \textit{i}=$68^{\circ}$ \\
\hline
0.063 & 0.95730 & 0.95799 & 0.95773 & 0.95836 \\
0.064 & 0.95794 & 0.95826 & 0.95800 & 0.95898 \\
0.066 & 0.95879 & 0.95876 & 0.95819 & 0.95915 \\
0.067 & 0.95938 & 0.95900 & 0.95912 & 0.95939 \\
0.069 & 0.95947 & 0.96014 & 0.95993 & 0.95986 \\
0.070 & 0.96036 & 0.96069 & 0.96014 & 0.95977 \\
0.071 & 0.96022 & 0.96056 & 0.96004 & 0.95999 \\

\hline
\end{tabular}
\end{center}
\end{table}

\end{document}